\documentclass[aps,floatfix,twocolumn,a4paper,showpacs, nofootinbib,superscriptaddress,10pt]{revtex4}
\usepackage{graphicx,float}\usepackage{graphicx,float}
\usepackage[all]{xy}
\usepackage{amsmath,upgreek}
\usepackage{amssymb}
\usepackage{color}
\usepackage{epsfig}		
\usepackage{graphicx,epstopdf}
\usepackage{subfigure}
\usepackage{pdfpages}
\usepackage[colorlinks,hyperindex]{hyperref}

\definecolor{green1}{RGB}{0,128,0} 
\hypersetup{hidelinks,backref=true,pagebackref=true,hyperindex=true,colorlinks=true,breaklinks=true,urlcolor= blue}
\hypersetup{%
 colorlinks = true,
 linkcolor = blue,
 citecolor = green1,
}
\usepackage{bookmark,textgreek}
\usepackage{hyperref,color,xcolor}
\hypersetup{hidelinks,hyperindex=true,colorlinks=true,breaklinks=true,urlcolor= blue}
\hypersetup{%
 colorlinks = true,
 linkcolor = blue
}

\newcommand{\bes}{\begin{subequations}}
\newcommand{\ees}{\end{subequations}}
\def\ben{\begin{eqnarray}}
\def\een{\end{eqnarray}}
\def\be{\begin{equation}}
\def\ee{\end{equation}}

\def\bb#1{\mbox{\footnotesize $(#1)$}}
\def\0{\mbox{\tiny $0$}}
\def\1{\mbox{\tiny $1$}}
\def\2{\mbox{\tiny $2$}}
\def\3{\mbox{\tiny $3$}}
\def\4{\mbox{\tiny $4$}}
\def\5{\mbox{\tiny $5$}}
\def\6{\mbox{\tiny $6$}}
\def\7{\mbox{\tiny $7$}}
\def\8{\mbox{\tiny $8$}}
\def\9{\mbox{\tiny $9$}}
\def\mi{\mbox{\tiny $-$}}

\begin{document}

\title{Cosmological comoving behavior of the configurational entropy}
\author{A. E. Bernardini} 
\affiliation{Departamento de F\'isica, Universidade Federal de S\~ao Carlos,
PO Box 676, 13565-905, S\~ao Carlos, SP, Brazil}
\email{alexeb@ufscar.br}
\author{R. da Rocha}
\email{roldao.rocha@ufabc.edu.br}
\affiliation{Federal University of ABC, Center of Mathematics, Computing and Cognition, 09210-580, Santo Andr\'e, Brazil}\email{roldao.rocha@ufabc.edu.br}

\begin{abstract}
It has been shown that a functional dependence of the configurational entropy (CE) density exhibits a cosmological comoving
behavior related to the scale parameter, $a$, of the FRW flat universe. 
Such a circumstantial comoving behavior shows that the CE evolves with $a^{-3}$ for fermionic and bosonic radiation and with $a^{-3/2}$ for non-relativistic fermionic matter. The results are discussed in the domain of matter and radiation, as well as in a simplified context of the $\Lambda$CDM cosmology. Our results suggest that the CE can work as a theoretical driver for particle and nuclear interactions in the context of large scale cosmic events.
\end{abstract}

\maketitle

\section{Introduction}

Since Boltzmann-Gibbs secular contributions to the interpretation of the thermodynamics have established the statistical entropy definition, entropic information measures have been considered in the quantification of the informational and organizational contents encoded by physical systems.
Once specialized to stochastic processes, such a statistical interpretation of the Nature was reconstructed in terms of the Shannon's information entropy paradigm which, through a lattice approach containing finite frequency modes, has been 
applied in a plethora of problems in Physics \cite{Witten:2018zva}.
From an informational viewpoint, the Shannon entropy is equivalent to the Boltzmann-Gibbs entropy in expressing the disorder level of a physical system \cite{shannon}.
More subtly, it gives rise to the concept of configurational entropy (CE) encoded by the energy density of physical solutions \cite{Gleiser:2011di,Gleiser:2012tu}.
Besides encompassing the informational content of the spatial complexity of a localized system \cite{Gleiser:2014ipa,Sowinski:2015cfa}, the CE is correlated to the configuration of wave modes in a physical system \cite{Gleiser:2011di,Gleiser:2012tu}. 
Through an analogy with the Shannon's lattice approach, it measures the logarithmic evaluation of the number of bits required to describe the organization of a system. 
In particular, the critical points of the CE have been identified as more dominant events -- in correspondence to all other modes -- in the sense that they are more frequently detected in Nature. 
Besides being relevant in the context of particle physics \cite{Bernardini:2016qit,Braga:2017fsb,Bernardini:2018,Ferreira:2019inu} and AdS/CFT theories \cite{Bernardini:2016hvx, Braga:2016wzx, Lee:2017ero}, phase transitions correlated to CE critical points have also been identified for several physical systems \cite{Gleiser:2014ipa,Sowinski:2017hdw,Sowinski:2015cfa,Gleiser:2018kbq}. 

Therefore, given the relevance of the CE as a driver for particle \cite{Bernardini:2016hvx,Bernardini:2016qit,Bernardini:2018,Colangelo:2018mrt} and even nuclear interactions \cite{Karapetyan:2018oye,Karapetyan:2018yhm,Karapetyan:2016fai,Karapetyan:2017edu}, one may suppose that it could exhibit a similar pattern at large scale cosmic events involving such particles. 
Hence, this article just describes the setup conditions to include the computation of the CE in the standard model cosmology, where the CE is realized as an alternative entropic quantifier for the wave modes encoded by the energy density of the components of the cosmic inventory.
The CE for fermionic and bosonic radiation and matter contributions are calculated in the context of the standard model cosmology, for the homogeneous Friedmann-Robertson-Walker (FRW) background flat universe. 
The results are specialized to radiation dominated (RD) and matter dominated (MD) regimes, and extended to a $\Lambda$CDM effective description. Unexpectedly, the results exhibit a comoving behavior which endorses the relevance of the CE in identifying some eventual covariant property of the information encoded by cosmological background fluids in the Universe.

This article is devised as follows.
Sect. \ref{eqnii} is devoted to a brief review of the CE theoretical grounds. 
A short view of the statistical mechanics that underlie the driving concept of conditional entropy, when extended to the continuum limit, is provided.
It reports about how collective coordinates, associated to the energy density of the system, allows one to establish a relationship between the conditional entropy, which supports the CE definition, and the thermodynamical entropy (cf. Ref.\cite{Bernardini:2016hvx}).
In Sect. \ref{eqniii}, the CE is computed for several components of the cosmic inventory in the context of the FRW flat universe.
The results are obtained and discussed in the domain of matter and radiation and through an approach for the $\Lambda$CDM cosmology.
Our conclusions are drawn in Sect. \ref{eqniv}.

\section{Configurational entropy in the elementary context}
\label{eqnii}

The Shannon information paradigm \cite{Gleiser:2011di}, which is specialized for discrete systems consisting of $n$ quantum modes, leads to a logarithmic quantifier for information given by $S = -{{\sum_{s=1}^n}} g_s\,\ln(g_s)$, the information entropy \cite{shannon}, where $g_s$ denotes the $s$-mode probability density function.
Generically, the recurrent (re)formulation, and even the generalization, of the Shannon concepts are very usual.
Hence, before pragmatically establishing the manipulative tool for computing the CE, one can notice, for instance, that
departing from the Gibbs entropy connected to thermodynamic ensemble of microstates,
${S=-k_{{B} }\sum_s \pi_{s}\ln \pi_{s}\,}$
where $\pi_i$ is the probability of measuring a microstate, $s$, and $k_B$ is the Boltzmann constant, the von Neumann entropy was defined as
${S=-k_{{B} }\,{\rm {Tr}}(\rho \ln \rho )\,}$, where $\rho = \vert\psi\rangle\langle \psi\vert$ is the density matrix associated to the quantum state, $\vert\psi \rangle$.
Likewise, depending on the supporting statistics and on the ensemble subtleties, the meaning of the Shannon entropy has already been re-formulated in terms of a generic expression, $-{{\sum_{s=1}^n}} g_s\,{\rm log}_b(g_s)$, in order to encompass some particular entropy definitions as, for instance, the Nat entropy and the Hartley entropy, respectively for $b=e$ and $b=10$ (besides $b = 2$ for the Shannon one).

In such an enlarged context, the proposition of a conditional entropy carrying out the information encoded by the energy density of a physical system, i.e., the CE, has been initially motivated by the fact that the continuum limit of the Shannon entropy rigorously lacks some substantial properties that underlie the definition of entropy (see Ref.\cite{Bernardini:2016hvx} and Refs. therein for a complete overview).

The CE, as an extension of the Shannon entropy, is thus constructed from the Fourier transform of the energy density, ${G}(x)$, i.e.
\begin{equation}
\mathcal{G}(k)=\frac{1}{{2\pi }}\int \;e^{ik x}{G}(x)\,dx,
\label{eqncollectivecoordinates}
\end{equation}
which defines an analogous continuum limit of the collective coordinates in statistical mechanics, as ${G}(x)\sim\sum e^{-ik_n x}\mathcal{G}(k_n)$. That is the discretized version of Eq. (\ref{eqncollectivecoordinates}) for periodic functions where $n$ corresponds to the index of modes in the physical system. 

The energy density is indeed suitable for defining the correspondent conditional entropy because it encodes more enhanced elements of the physical system. Of course, it is much more evident for spatially localized structures, as those described by lumps and kinks, and by an infinite set of one dimensional (non)topological defects \cite{Braga:2018fyc,daSilva:2017jay,Ma:2018wtw,Gleiser:2013mga,Gleiser:2015rwa,Correa:2015ako,Correa:2016pgr,Bazeia:2018uyg,Alves:2014ksa}. 
For the energy density Fourier decomposed into components weighted by the modes $k_n$, the normalized structure factor is constructed upon the collective coordinates as \cite{Gleiser:2011di,Gleiser:2012tu,Bernardini:2016hvx}
\begin{equation}
g(k_n)=\frac{\langle\;\left\vert \mathcal{G}(k_n)\right\vert ^{2}\;\rangle}{\;\sum_{j=1}^n\langle\;\left\vert \mathcal{G}(k_n)\right\vert ^{2} \rangle\,}.
\label{eqncollective1}
\end{equation}
The structure factor, $g$, measures the relative weight carried by the corresponding mode $k_n$.

For the continuous limit, with Eq.~\eqref{eqncollectivecoordinates} extend to $\mathbb{R}^n$, the modal fraction 
\cite{Gleiser:2012tu,Sowinski:2015cfa} is written as
\begin{eqnarray}
\upepsilon(\mathbf{k}) = \frac{|\mathcal{G}(\mathbf{k})|^{2}}{ \int_{\mathbb{R}^n} |\mathcal{G}(\mathbf{k})|^{2}d^n \mathbf{k}},\label{modalf}
\end{eqnarray}
which gives a correlation probability distribution that quantifies how much a $n$-dim $\mathbf{k}$ wave mode contributes to the power spectrum associated to the energy density.
Therefore, the CE measures the information content of the spatial profile that characterizes the energy density, $\mathcal{G}(\mathbf{x})$, by means of an association with a probability density, through the decomposition of the probability amplitude in terms of the Fourier wave modes. 
The CE results into an expression given by \cite{Gleiser:2012tu,Sowinski:2015cfa}
\begin{eqnarray}
S^{\mbox{\tiny{CE}}}[\mathcal{G}] = - \int_{\mathbb{R}^n}{\upepsilon_\diamond}(\mathbf{k})\log {\upepsilon_\diamond}(\mathbf{k})\, d^n\mathbf{k}\,,
\label{confige}
\end{eqnarray}
for $\upepsilon_\diamond(\mathbf{k})=\upepsilon(\mathbf{k})/\upepsilon_{\rm max}(\mathbf{k})$.
$S^{\mbox{\tiny{CE}}}[\mathcal{G}]$ defines the best lossless compression of any exchange of information which involves any interface inside/outside the system. It increases with the unpredictability of the information content of the system.
As preliminarily discussed, the Shannon entropy discretized approach is re-interpreted as an entropy of spatial profile: an informational measure of the complexity of a system that has some localized shape \cite{Gleiser:2011di,Gleiser:2013mga,Gleiser:2014ipa}.
The less information, the lower the information entropy encoded by the frequency modes required to represent the spatial profile of the energy density\cite{Gleiser:2011di,Gleiser:2013mga}.

To extend the meaning of the CE to the context of the standard model cosmology, one should correctly interpret the modal decomposition provided by the statistical distributions that drive cosmological energy distributions. 
For the {\em integrated energy density} of an $i$-component of the cosmic inventory, $\varrho_i$, equivalent to a rate of energy by volume, a straightforward identification with the modal fraction from Eq.~(\ref{eqncollectivecoordinates}) is given by 
\begin{eqnarray}\label{diemm}
\varrho_i \propto \int_{\mathbb{R}^n} |\mathcal{G}(\mathbf{k})|^{2} d^n \mathbf{k}.
\end{eqnarray}
Therefore, the above-mentioned localization aspect is only evinced from the profile of the momentum distribution functions for cosmological fermion and boson(photon) backgrounds. 
Moreover, the assumption/verification of the existence of a critical density, $\varrho_C$, close to unity, so that $\sum_i \Omega_i = 1$, with $\Omega_i=\varrho_i/\varrho_C$, naturally suggests a normalized probability interpretation to such cosmological background energy density components, $\Omega_i$.
Given that $\varrho_i$ is identified by Eq.~\eqref{diemm}, it turns the CE, $S^{\mbox{\tiny{CE}}}[\mathcal{G}]$, into a natural choice for measuring the informational content of the cosmological background.

\section{Configurational entropy in the background cosmic inventory}
\label{eqniii}

For the homogeneous FRW flat universe with energy density, $\varrho\bb{\tau}$, one has the driving Friedmann equation for the scale factor, $a\bb{\tau}$,
\begin{eqnarray}
 \left( {\dot{a}\over a} \right)^{\2} &=& {8\pi\over3}G a^{\2} {\varrho} \,,
\label{eqnfriedmann01}
\end{eqnarray}
where {\em dots} denote conformal (dimensionless) time derivatives, $d/d\tau$.
The dynamical evolution of ${\varrho}$ depends on its dependence on $a$, ${\varrho}\equiv {\varrho}\bb{a}$, which can be driven by some state equation or even by the intrinsic thermodynamic properties of some associated statistical ensemble.
Considering the latter hypothesis, one can introduce the phase-space distribution of particles given by
\begin{equation}
f\bb{p\bb{E}, \mathbf{x}, \tau}\,dx^{\1} dx^{\2} dx^{\3} dp_{\1} dp_{\2} dp_{\3}/(2\pi)^3 = dN,
\end{equation}
where $f$ is a Lorentz scalar that is also invariant under canonical transformations, $dN$ denotes the number of particles in an infinitesimal phase-space volume $dx^{\1} dx^{\2} dx^{\3} dp_{\1} dp_{\2} dp_{\3}$, and it has been assumed that $\hbar =1$.
Assuming a comoving behavior of the background temperature parameterized by $T\equiv T\bb{\tau} \equiv T\bb{a}$, the reduced form,
$f\bb{p\bb{E, \tau}}$,
can be identified either with the Fermi-Dirac distribution, $f_+$, for fermions, or with the Bose-Einstein distribution, $f_-$, for bosons,
\begin{equation}
f_{\pm}\bb{E, \tau} = \sigma \, \left(\exp{\left[\frac{E}{T} - \frac{\mu_{\0}}{T{\0}}\right]} \pm 1\right)^{\mi\1},
\label{eqndist1}
\end{equation}
where the factor $\sigma$ denotes de spin degrees of freedom, and it has been assumed that $k_B =1$, and that $T_{\0} = \mu_{\0} (T/\mu)$ corresponds to the background temperature today.
Even if the thermodynamics of a test fluid is constrained by a chemical potential, $\mu\sim \mu\bb{a}$, its contribution to the distribution function is often discarded by setting $\mu = 0$.
The trick to turn the choice of $\mu = 0$ into a generic one is performed through a simple manipulation involving the Boltzmann equation \cite{Ma94,Meu01,Meu02,Meu03}: the condition
${\mbox{d} f}/{\mbox{d} \tau} ={\mbox{d} f}/{\mbox{d} a}= 0$ is implemented through the assumption that $\mu/T = \mu_{\0}/T_{\0}$ does not depend on $a$ and, therefore, the chemical potential follows the same temperature fluctuations on time.
Consequently it does not contribute to first-order corrections of the Boltzmann equation and the following constraint between $f$ and $T$ is maintained,
\begin{equation}
\frac{\partial f}{\partial E} = \frac{E}{p} \frac{\partial f}{\partial p}
 = - \frac{1}{E} \left(T \frac{\partial f}{\partial T}\right).
\label{eqndist2}
\end{equation}
It provides a simplified realistic condition to depict (semi)analytical expressions for the transitory regime of a cosmological fluid, from relativistic to non-relativistic dynamics. For ultra-relativistic (non-massive) fluids, one thus has
\begin{eqnarray}
\left(\sigma_{\gamma}^{\mi\1}\right)\varrho_{\gamma}
&=&\int \frac{{d}^{^{\3}}\hspace{ -.1 cm}\mathbf{p}}{(2\pi)^{\3}} \frac{1}{e^{p/T} - 1} p
\nonumber\\
&=& \frac{4\pi}{(2\pi)^{\3}} T^{\4} \int_{0}^{\infty} \frac{y^{\3}~{d}y}{e^y - 1} \equiv\frac{\pi^{\2}}{30} T^{\4},
\label{rhophotons}
\end{eqnarray}
and
\begin{eqnarray}
\left(\sigma_{\nu}^{\mi\1}\right)\varrho_{\nu} 
&=& \int \frac{{d}^{^{\3}}\hspace{ -.1 cm}\mathbf{p}}{(2\pi)^{\3}} \frac{1}{e^{p/T} + 1} p
\nonumber\\ &=&
 \frac{4\pi}{(2\pi)^{\3}} T^{\4} \int_{0}^{\infty} \frac{y^{\3}~{d}y}{e^y + 1} \equiv\frac{7\pi^{\2}}{240} T^{\4},
\label{rhofermions}
\end{eqnarray}
respectively for fermion ($\nu$) and photon ($\gamma$) energy densities, where $\sigma_i$ shall be absorbed by the normalization constraints along the computation of $S^{\mbox{\tiny{CE}}}[\mathcal{G}]$.
Similarly, for the non-relativistic fermionic component, one has the semi-analytical expression,
\begin{eqnarray}
\left(\sigma_{\mathcal{M}}^{\mi\1}\right)\varrho_{\mathcal{M}} 
&=& \int \frac{{d}^{^{\3}}\hspace{ -.1 cm}\mathbf{p}}{(2\pi)^{\3}} \frac{1}{e^{E/T} + 1} E
\nonumber\\ &=&
 \frac{4\pi}{(2\pi)^{\3}} T^{\4} \int_{0}^{\infty} \frac{y^{\2}\sqrt{\mathcal{M}^{\2}+y^{\2}}~{d}y}{e^{\sqrt{\mathcal{M}^{\2}+y^{\2}}} + 1}
\label{rhofermions2}
\end{eqnarray}
where $\mathcal{M}$ is the mass of the relativistic particle given in units of $k_B T_{\0}$. 

\begin{figure}
\includegraphics[scale=0.66]{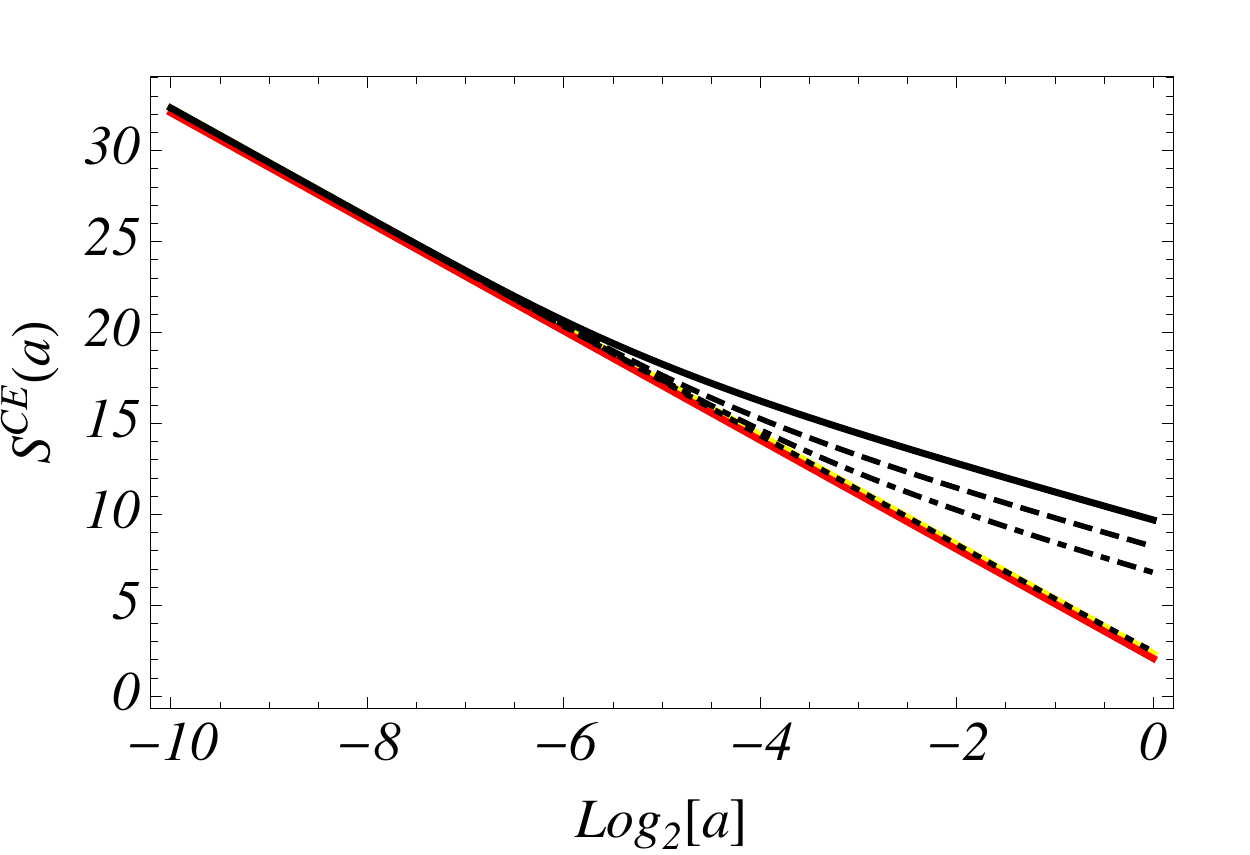}
\renewcommand{\baselinestretch}{.85}
\caption{\footnotesize{
(Color online) Configurational entropy as function of the scale factor (in logarithmic scale) for ultrarelativistic fermions (red solid line) and photons (yellow solid line) and for non-relativistic fluids of fermionic particles with $\mathcal{M} = 2k_B T_{\0}$ (black dotted line),\, $50k_B T_{\0}$ (black dashed-dotted line),\, $100k_B T_{\0}$ (black dashed line) and $200k_B T_{\0}$ (black solid line).
}}
\label{fig01}
\end{figure}

Finally, by following the identification from Eq.~(\ref{diemm}), after numerical integrations, the CE is depicted in Fig.~\ref{fig01} (in logarithmic scale) for ultrarelativistic fermions and photons (red and yellow solid lines) and for non-relativistic fluids of fermionic particles with $\mathcal{M} = 2k_B T_{\0},\, 50k_B T_{\0},\, 100k_B T_{\0}$ and $200k_B T_{\0}$.

\begin{figure}
\includegraphics[scale=0.66]{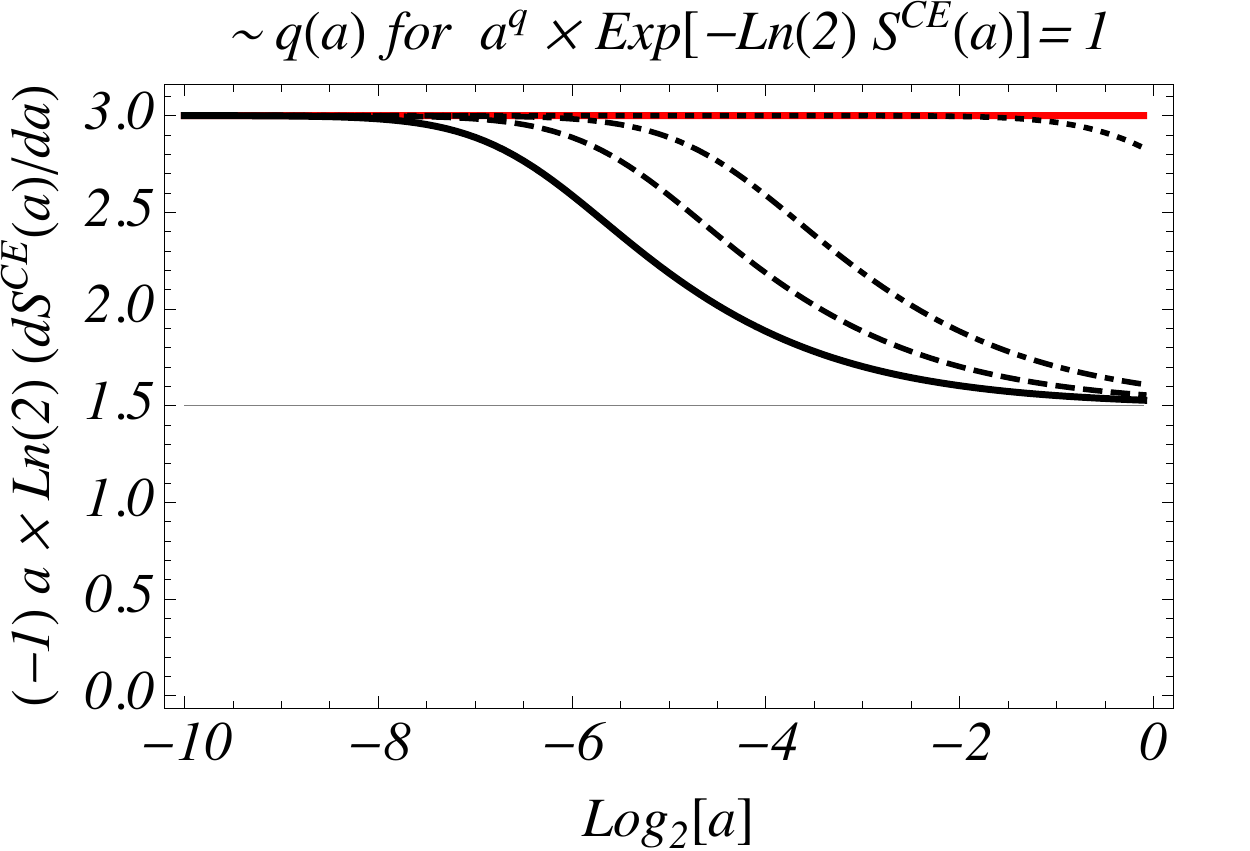}
\renewcommand{\baselinestretch}{.85}
\caption{\footnotesize{
(Color online) Derivative of the configurational entropy, $dS^{\mbox{\tiny{CE}}}/d \mbox{log}_2\bb{a}$ and its corresponding comoving behavior. The plot data (line style) are in correspondence with Fig.~\ref{fig01}.}}
\label{fig02}
\end{figure}

The logarithmic scale clears up the comoving behavior of $S^{\mbox{\tiny{CE}}}$ in the ultra-relativistic and non-relativistic limits. Such a non expected comoving behavior shows that CE evolves as an approximated functional intrinsically given by $a^{q\bb{a}} \exp\left[-\ln(2)\,S^{\mbox{\tiny{CE}}}\bb{a}\right]$ with $q\bb{a} = {3}$ for fermionic and bosonic radiation, and with $q = 3/2$ for non-relativistic fermionic matter. 
For the extreme of ultra- and non-relativistic regimes, the results for $q$ are exact and correspond to the values of the logarithmic derivative of $S^{\mbox{\tiny{CE}}}$, $dS^{\mbox{\tiny{CE}}}/d \mbox{log}_2\bb{a}$, as it can be depicted in Fig.~\ref{fig02}.

Turning back to the scale factor dependence on the conformal time, $a\bb{\tau}$, the solutions of Eq.~\eqref{eqnfriedmann01} for RD($\gamma$) and MD($\mathcal{M}$) universes, and for the cosmological constant regime ($\Lambda$), with the respective equations of state given by $\mathcal{P}_{\gamma} = \varrho_{\gamma}/3$, $\mathcal{P}_{\mathcal{M}} = 0$ and $\mathcal{ P}_\Lambda = -\varrho_\Lambda$, are given by \cite{Meu01}
\begin{equation}
\label{rho(a)}
 \varrho_{\gamma} = \varrho_0 \frac{\Omega_{\gamma}}{a^4}\,,
\quad
 \varrho_{\mathcal{M}} = \varrho_0 \frac{\Omega_{\mathcal{M}}}{a^3}\,,
\quad
\mbox{and} \quad
 \varrho_\Lambda=\varrho_0\, \Omega_\Lambda\,,
\end{equation}
where $\varrho_0\equiv \varrho_C$ is the present (integrated) Universe's energy density, and from which the expansion factor scales as $a\propto\tau$ in the RD era and as $a\propto\tau^{\2}$ in the MD era.
Through an effective simplified analysis, departing from the RD era parameterized by
\begin{equation}
a_{\gamma}\bb{\tau} \sim \Omega_{\gamma}^{\1/\2} \, \tau,
\end{equation}
and consistently following the analytical continuity conditions, extensions to MD and $\Lambda$-dominated ($\Lambda$D) background scenarios are straightforwardly obtained as
\begin{equation}
a_{\mathcal{M}}\bb{\tau} \sim \frac{\Omega_{\mathcal{M}}}{4}\left( \tau + \frac{\Omega_{\gamma}^{\1/\2}}{\Omega_{\mathcal{M}}}\right)^{\2},
\end{equation}
and
\begin{equation}
a_{\Lambda}\bb{\tau} \sim \frac{1}{\Omega_\Lambda^{\1/\2}}
\left(\frac{3}{\Omega_{\mathcal{M}}^{\1/\3}\,\Omega_{\Lambda}^{\1/\6}} - \frac{\Omega_{\gamma}^{\1/\2}}{\Omega_{\mathcal{M}}}-\tau\right)^{\mi\1},
\end{equation}
respectively for the MD and $\Lambda$D eras, where ${3/ 8\pi G}= \varrho_C$ has been set equal to unity, and $\Omega_i = \varrho_i/\varrho_0$, with $i = \gamma,\, \mathcal{M},\, \Lambda$\footnote{An alternative approach for radiation-to-matter transition eras is given by \cite{Meu02}
\begin{equation}
 a(\tau)\approx 4\Omega_{\mathcal{M}}\tau^2 + \Omega_{\gamma}^{\1/\2}\tau \,,
\label{a(eta)}
\end{equation}
and during the $\Lambda$-dominated era by
\begin{equation}
a(\tau)\approx\left[3\left(\frac{\Omega_\Lambda^{\1/\3}}{\Omega_{\mathcal{M}}^{\1/\3}}-\frac{4{\Omega^{\1/\2}_{\gamma}\Omega^{\1/\2}_\Lambda}}{\Omega_{\mathcal{M}}}\right)-\tau\right]^{\mi\1}\,.
\label{a(eta)2}
\end{equation}
However, it does not modify the qualitative interpretation of our results.}.

\begin{figure}
\includegraphics[scale=0.39]{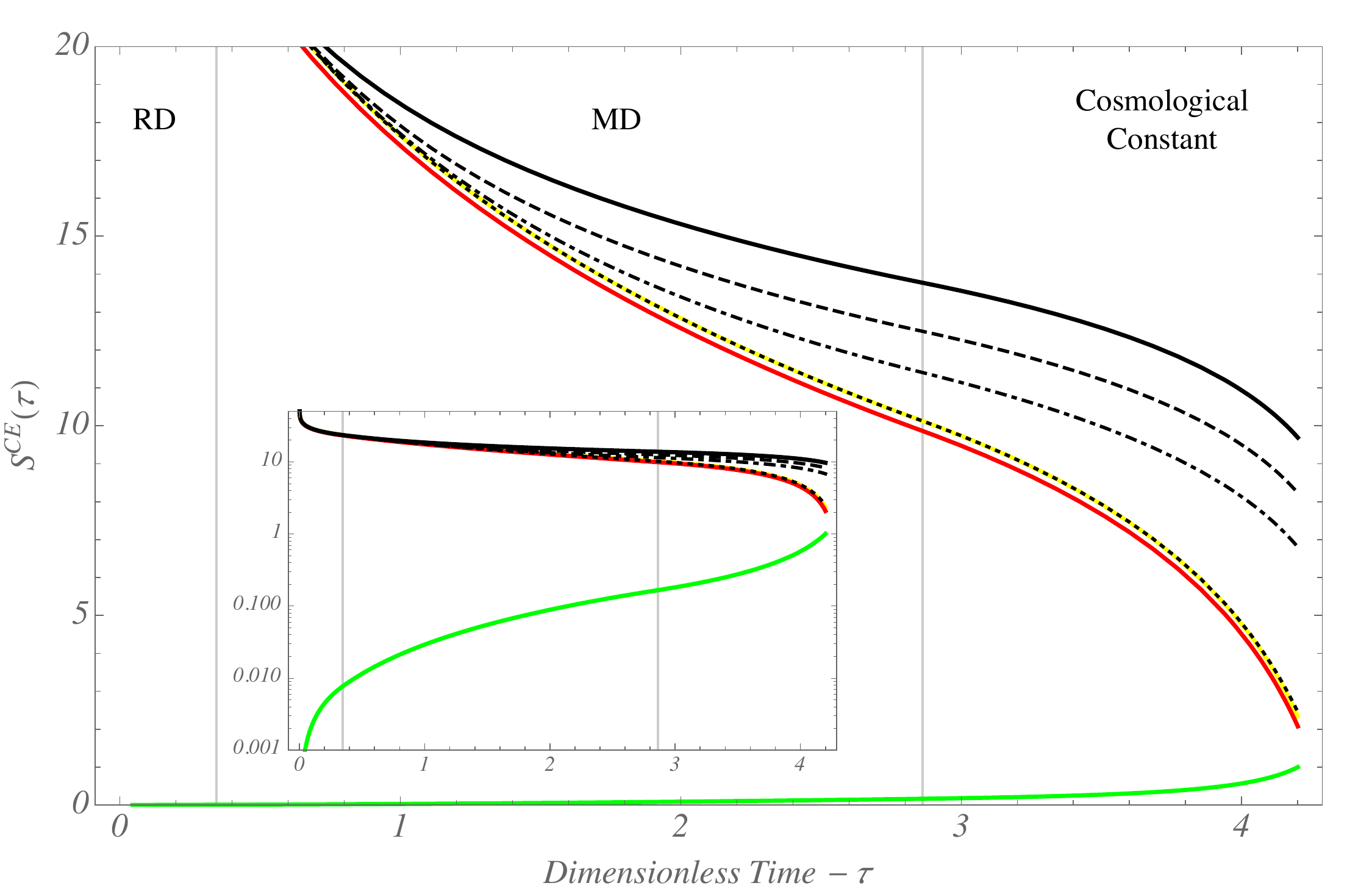}
\renewcommand{\baselinestretch}{.85}
\caption{\footnotesize{
(Color online) Configurational entropy as function of the comoving time, $\tau$. The plot data (line style) are in correspondence with Fig.~\ref{fig01}. One can notice an omnipresent decreasing regime which is negatively accelerated at RD and MD eras, and positively accelerated at the cosmological constant phase.}}
\label{fig03}
\end{figure}

In this case, the conformal time for the $50\%-50\%$ rate between radiation and matter densities, $\tau_{eq}$, for the $50\%-50\%$ rate between matter plus radiation and cosmological constant $\Lambda$ densities, $\tau_{\mathcal{M}\Lambda}$, as well as the conformal time for the present $\Lambda$ dominance at $a=a_{\0} = 1$, $\tau_{\Lambda 0}$, are described in terms of the density parameters through the recurrence relations given by
\begin{eqnarray}
\tau_{eq} &\sim& \frac{\Omega_{\gamma}^{\1/\2}}{\Omega_{\mathcal{M}}}, \nonumber\\
\tau_{\mathcal{M}\Lambda} &\sim& \frac{2}{\Omega_{\mathcal{M}}^{\1/\3}\,\Omega_{\Lambda}^{\1/\6}} -\tau_{eq} , \nonumber\\
\tau_{\Lambda 0} &\sim& \frac{3}{2}\tau_{\mathcal{M}} +\frac{1}{2}\tau_{eq} -\frac{1}{\Omega_\Lambda^{\1/\2}}.
\label{eqntimes}
\end{eqnarray}
In order to set a realistic correspondence with the phenomenology, it has been adopted the values of $\Omega_{\gamma} \sim 0.01$ for the sum of photon and fermionic (neutrino) radiation, $\Omega_{\gamma} \sim 0.29$ for cold matter, and $\Omega_{\Lambda} \sim 0.70$ for dark energy\footnote{Fluctuations about such values should not qualitatively affect our results.}.
The results for the comoving time dependence of the CE, for radiation, matter and cosmological constant dominated eras, are depicted in Fig.~\ref{fig03}, from which one can notice that, as a consequence of the comoving correspondence, the CE follows opposite acceleration regimes when compared with the expansion rate of the Universe (green solid line). At late times, the suppression of the informational content is much more drastic for ultra-relativistic fluids. It means that, from the informational paradigm viewpoint, cold dark matter represent an enormous and much more relevant contribution to the CE of the Universe.

\section{Brief Conclusions}
\label{eqniv}

The phenomenology supported by the FRW background cosmology has been probed along the last decades in order to quantitatively  ascertain the predictions for the elementary properties of the Universe: its accelerated expansion behavior, the space-time flatness, the cosmic inventory composition, the matter and radiation perturbation pattern, the gravitational wave spectrum, and also an additional set of complementary information \cite{Dodelson,Ma94}.
This corresponding background behavior is systematically supposed to evolve isentropically, once a classical definition of entropy density is identified by $s = (\epsilon + P) /T $ where $\epsilon$ and $P$ are respectively energy and pressure background densities, and $T$ is the temperature.
In fact, in the standard cosmological model framework, fluid-like components and homogeneous fields, once embedded into the cosmic inventory, evolve all together isentropically. Generically, non-isentropic fluids, where $dS/da \neq 0$, with $S = s\,a^{3}$, demand for the inclusion of compensating components into the driving Lagrangian density as to recover the entropy comoving behavior, i. e. $dS/da = 0$.
Therefore, as a preliminary assumption, any alternative consistent definition for quantifiers of the information content of the Universe should at least obey a similar comoving behavior, even if it is segmented to each background dominating era.
It avoids, for instance, the inclusion of spurious compensation mechanisms -- due to field dynamics or even due to some mass varying mechanism -- for entropy (information content) variations.
For all these reasons, it is natural to expect the identification of regimes where functions of $S^{\mbox{\tiny{CE}}}$ follow constant patterns, as to reflect the comoving behavior of the CE through some {\em iso(C)entropic} behavior like $d(a^n\,e^{- \ln(2)\,S^{\mbox{\tiny{CE}}}})/da = 0$ in the FRW cosmology.

To implement such a hypothesis, the CE was computed for several components of the cosmic inventory in the framework of FRW flat cosmology, in the domain of matter and radiation and through an approach for the domain of the cosmological constant. It has been shown that a functional dependence of the configurational entropy (CE) density exhibits such a cosmological comoving behavior related to the scale parameter, $a$. That circumstantial comoving behavior shows that the CE density evolves with $a^{-3}$ for fermionic and bosonic radiation and with $a^{-3/2}$ for non-relativistic fermionic matter. 

It is relevant to emphasize that, in case of the standard model cosmology, the interpretation of complexity and localization aspects provided by the implementation of the CE were evinced from the profile of the momentum distribution functions for cosmological fermion and boson(photon) backgrounds. The existence of a critical density, $\varrho_C$, close to unity, so that $\sum_i \Omega_i = 1$, with $\Omega_i=\varrho_i/\varrho_C$, where the values of $i$ designate the components of the cosmic inventory, inputs the statistical meaning of normalized probabilities to the cosmological background energy density components, $\varrho_i$. It helpfully supports the suggestion of the CE as a natural choice for measuring the informational content of the Universe, from which, its possible to assert that, at the present epoch, cold dark matter gives a much more relevant contribution to the informational content of the Universe.

To conclude, it is important to notice that our entire analysis was performed in terms of dimensionless quantities related to standard model cosmological parameters. Any eventual change in the phenomenological cosmology can be fitted onto the results present here. Of course, after the setup idea, the role of the CE in particle physics cosmology can also be explored in the context of cosmic perturbations as they indeed correspond to spatially localized distributions of energy.

\paragraph*{Acknowledgments:} The work of AEB is supported by the Brazilian Agencies FAPESP (grant 2018/03960-9) and CNPq (grant 300831/2016-1). RdR~is grateful to FAPESP (Grant No. 2017/18897-8) and to the National Council for Scientific and Technological Development -- CNPq (Grant No. 303293/2015-2), for partial financial support.

\newpage{\pagestyle{empty}\cleardoublepage}

\end{document}